\documentclass[conference]{IEEEtran}
\IEEEoverridecommandlockouts

\usepackage{cite}
\usepackage{amsmath,amssymb,amsfonts}
\usepackage{algorithmic}
\usepackage{graphicx}
\usepackage{textcomp}
\usepackage{xcolor}
\def\BibTeX{{\rm B\kern-.05em{\sc i\kern-.025em b}\kern-.08em
    T\kern-.1667em\lower.7ex\hbox{E}\kern-.125emX}}
\begin{document}

\title{Enhancing LLM-based ASR Accuracy with Retrieval-Augmented Generation\\
}


\author{\IEEEauthorblockN{Shaojun Li}
\IEEEauthorblockA{\textit{Huawei TSC} \\
Beijing, China \\
lishaojun18@huawei.com}
\and
\IEEEauthorblockN{Hengchao Shang}
\IEEEauthorblockA{\textit{Huawei TSC} \\
Beijing, China  \\
shanghengchao@huawei.com}
\and
\IEEEauthorblockN{Daimeng Wei}
\IEEEauthorblockA{\textit{Huawei TSC} \\
Beijing, China \\
weidaimeng@huawei.com}
\and
\IEEEauthorblockN{Jiaxin Guo}
\IEEEauthorblockA{\textit{Huawei TSC} \\
Beijing, China \\
guojiaxin1@huawei.com}
\and
\IEEEauthorblockN{Zongyao Li}
\IEEEauthorblockA{\textit{Huawei TSC} \\
Beijing, China \\
lizongyao@huawei.com	}
\and
\IEEEauthorblockN{Xianghui He}
\IEEEauthorblockA{\textit{Huawei TSC} \\
Beijing, China \\
hexianghui@huawei.com}
\and
\IEEEauthorblockN{Min Zhang}
\IEEEauthorblockA{\textit{Huawei TSC} \\
Beijing, China \\
zhangmin186@huawei.com}
\and
\IEEEauthorblockN{Hao Yang}
\IEEEauthorblockA{\textit{Huawei TSC} \\
Beijing, China \\
yanghao30@huawei.com}
}

\maketitle

\begin{abstract}
Recent advancements in integrating speech information into large language models (LLMs) have significantly improved automatic speech recognition (ASR) accuracy. However, existing methods often constrained by the capabilities of the speech encoders under varied acoustic conditions, such as accents. To address this, we propose LA-RAG, a novel Retrieval-Augmented Generation (RAG) paradigm for LLM-based ASR. LA-RAG leverages fine-grained token-level speech datastores and a speech-to-speech retrieval mechanism to enhance ASR accuracy via LLM in-context learning (ICL) capabilities. Experiments on Mandarin and various Chinese dialect datasets demonstrate significant improvements in ASR accuracy compared to existing methods, validating the effectiveness of our approach, especially in handling accent variations.

\end{abstract}


\begin{figure*}[t]
    \vspace{-0.5cm} %
    \small
    \centering
    \includegraphics[width = 12cm]{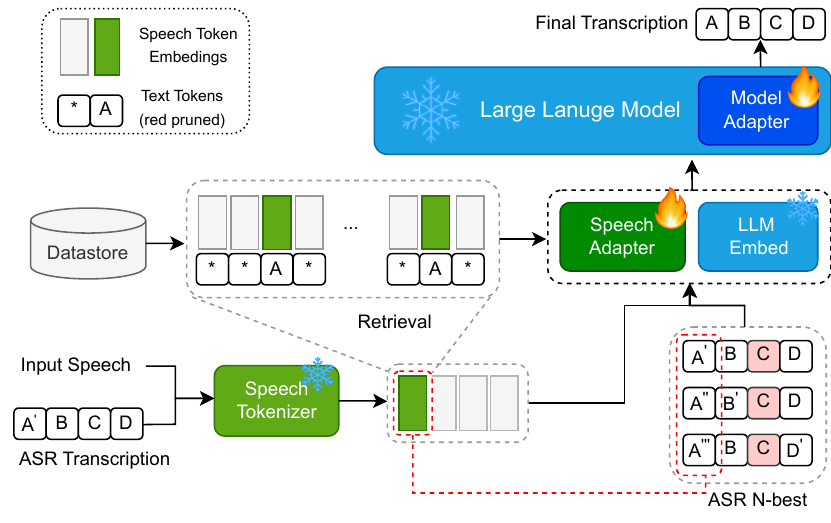}
    \vspace{-0.3cm} %
    \caption{Overview of proposed LA-RAG, The speech tokenizer is employed to generate aligned speech tokens and text tokens. With the 1th token as an example, the input of A' represents an incorrect token, with the corresponding speech token indicated in green, which is one of retention of N-best pruning. This speech token is subsequently used to query the datastore. The retrieval examples include the mappings between speech token and the correct token A. Ultimately, the examples, the input speech tokens and the N-best results, are transmitted to the LLM prompt for ICL via the adapter and embed process.}
    \label{fig:arch}
    \vspace{-0.5cm} %
\end{figure*}

\section{Introduction}
In recent years, there has been growing interest in integrating speech information into LLMs \cite{HyPoradise,speech_llama,whispering_llama}. These models have demonstrated remarkable efficacy in ICL capabilities to improve the ASR accuracy (LLM-based ASR). Initial studies typically input pure textual transcriptions into the LLM, often combining the ASR N-best results with instructions to prompt the LLM for error correction \cite{HyPoradise,Yang_2023,ma2023generativelargelanguagemodels}. In these studies, the LLM primarily serves as a text reranker or token selector. Concurrently, other studies have attempted to integrate pre-trained ASR models (most commonly using the speech encoder part) into LLMs with a modality adapter, such as Q-former, attention, or a projection to align the speech feature space with the textual space of the LLM \cite{speech_llama,yu2023connectingspeechencoderlarge,li2024usinglargelanguagemodel}. These approaches generally show improvements by leveraging rich acoustic signals. Further research has combined N-best results with speech encoders and even added denoising information \cite{whispering_llama,fathullah2023promptinglargelanguagemodels,chen2024itslatefusingacoustic,robust}. Such multi-source information integration usually leads to better performance. However, the performance ceiling of these methods is often limited by the capabilities of speech encoders. This is particularly evident when there is an acoustic feature mismatch between the training and test data of the speech encoder, such as in scenarios with accents where the encoder is insufficiently trained and the correct tokens do not appear in the N-best transcriptions. These methods struggle under such conditions. Usually, for traditional ASR models, domain adaptation or speaker adaptation can be used to address the issue of insufficient training \cite{Huang_Ye_Li_Gong_2021,10389732}. However, for LLM-based ASR, aside from the costly fine-tuning, this can be achieved through Retrieval-Augmented Generation (RAG) \cite{rag1,rag2}, allowing the LLM to learn external knowledge during inference.

Compared to token-level or semantic-level matching in text-based RAG, the challenge of RAG in LLM-based ASR stems from how to accurately retrieve relevant speech examples and how to prompt LLMs from inherently high sampling rate acoustic data. \cite{speech_icl} explores and proposes a speech LLM capable of performing unseen classification tasks for the first time. COSMIC \cite{COSMIC} pioneered this capability in more complex ASR tasks, showing significant ASR accuracy gains in context-biased tasks. However, the above methods only use random sampling for example selection and lack exploration of how to retrieve more similar examples. \cite{speech_rag} first explored RAG in LLM-based ASR and created a retrieval datastore. \cite{yang24b_interspeech} proposed using RAG to enhance SLU task. However, they only focused on entity retrieval or only used coarse-grained speech retrieval, which makes accurate speech matching difficult.

The construction of a fine-grained speech datastore for the LLM-based ASR task is hindered by a lack of precise speech-transcript alignment and the enormous volume of frame-level entries. Recently, in the speech retrieval augmentation task for small models, \cite{knn_ctc} and \cite{speaker_smoth} separately used Connectionist Temporal Classification (CTC) and Attention Encoder-Decoder (AED) pre-trained ASR models as speech tokenizers to force-align the speech features and text tokens. They established key-value pair mappings between speech features and text transcription tokens and retrieved the keys for each decoding step with a query extracted from hidden states, achieving effective performance. However, due to the large number of LLM parameters, the speed and storage consumption would be enormous if directly applied to LLM-based ASR.

Therefore, we propose a new \textbf{L}LM-based \textbf{A}SR \textbf{RAG} (\textbf{LA-RAG}) paradigm utilizing the above speech tokenizers, fully leveraging the LLM's ICL capabilities. Specifically, in the database creation phase, speech tokenizers are used to obtain token-level precise alignment knowledge between speech hidden states and golden transcription tokens as key-value pairs, and the mapping between each key-value pair and its whole sequence is also stored as a speech inverted index. In the generation phase, the ASR transcription is used to perform the same speech tokenizing on the input speech, and each speech token obtained is used to query the index. By grouping and filtering policies, similar examples at the sequence level are obtained. In addition, to reduce the learning burden on the model, a pruning policy is added to remove tokens with low error probability. Finally, we input the speech and its golden transcription example pairs, together with the input speech tokens and N-best transcriptions, as prompts into the LLM. Here, we introduce a speech adapter to align speech and text spaces, and a model adapter to learn the mapping relationship of speech tokens to the correct text tokens. Experiments on Mandarin and various Chinese dialect datasets demonstrate significant improvements in ASR accuracy compared to existing methods, especially in handling accent variations.

Our contributions are as follows:
\begin{itemize}
    \item We propose a fine-grained retrieval method for speech-to-speech, implemented using a pre-trained ASR model through a simple forced alignment technique.
    \item We introduce a novel RAG paradigm for LLM-based ASR. By enabling the LLM to learn the mapping relationship between speech tokens and text tokens.
    \item We apply these methods to LLM-based ASR, leading to a significant enhancement in the accuracy of ASR results.
\end{itemize}

\begin{table*}[]
\centering
\caption{CER (\%) of our LA-RAG compare to others on Mandarin and various Chinese dialect datasets} 
\label{main}
\begin{tabular}{@{}c|c|c|ccccccc@{}}
\hline
& w/ Datastore & w/ LLM & AISHELL & Mandarin & JiangHuai & JiLu  & ZhongYuan & Southwestern &Avg.   \\ \hline
Base ASR            &$\times$ &$\times$ &5.18	&12.18	&43.94	&31.61	&34.01	&31.42	&26.39   \\  \hline
HyPoradise          &$\times$ &$\checkmark$ &4.91	&12.1	&43.57	&30.97	&33.98	&31.33	&26.14    \\
Whispering LLaMA    &$\times$ &$\checkmark$ &4.69	&11.93	&43.02	&30.88	&33.53	&31.07	&25.85    \\ \hline
$k$NN-CTC           &$\checkmark$ &$\times$ &4.83	&12	    &43.41	&30.71	&32.6	&30.63	&25.70    \\
LA-RAG$_{CTC}$      &$\checkmark$ &$\checkmark$   &\textbf{4.56}	&11.86	&\textbf{41.8}	&\textbf{30.39}	&\textbf{31.96}	&29.6 &\textbf{25.03}   \\ 
LA-RAG$_{AED}$      &$\checkmark$ &$\checkmark$ &4.61	&\textbf{11.69}	&42.11	&30.65	&32.25	&\textbf{29.56}	&25.15       \\ \hline
Datastore size (Million Tokens) & - & - &38.4	&12.7	&0.9	&1.1	&1.6	&1.4	&9.35 \\ \hline


\end{tabular}
\vspace{-0.5cm} %

\end{table*}

\section{Method}
As shown in Figure \ref{fig:arch}, we leverage RAG for LLM-based ASR, to Enhancing ASR transcript accuracy. Our method includes four main parts: speech tokenizer, datastore creation, speech retrieval and LLM prompt.
\subsection{Speech Tokenizer}
Given speech transcription pair $(x,y)$,  we can extract the intermediate representations of $X$, denoted as $f(x)$, by a pre-trained AED/CTC model. For simplify, we use the output of the final encoder(for CTC)/decoder(for AED) layer's feed-forward network (FFN) as our speech token.
To be specific, for CTC model, improve from \cite{knn_ctc}, we use a more precise algorithm for forced alignment, described in \cite{K_rzinger_2020}, by generate a trellis matrix which represents the probability of labels aligned at time step and find the most likely path from the trellis matrix. Then, we can get each speech token $f(x_t)$ from $f_{CTC}(x,y)$ for each text token by remove the blank ones. For AED model, following \cite{speaker_smoth}, which can generates the context representation $f_{AED}(x,y_{<t})$ at each time step $t$ also as a speech token $f(x_t)$ for each text token. 

\subsection{Datastore Creation}
\label{sec:datastore}
For datastore creation, we utilize a speech tokenizer on each training data $(x,y)\in\mathcal{S}$. This process yields speech tokenizer, we get the speech token representation $f(x_t)$ as the key $k_t$ and the CTC/AED ground-truth label $y_t$ as the value $v_t$, creating a speech-text key-value pair $(k_t, v_t)$ for the $t$-th token. Additionally, the corresponding sequence $(f(x), y)$ for each key-value pair is also saved and will serve as a final prompt example for the LLM, providing richer contextual information. Extending this process across the entirety of the training set $\mathcal{S}$, we construct a datastore $\mathcal{(K,V,X,Y)}$ composed of token-level key-value pairs and their corresponding sequences. 
\begin{equation}
    (\mathcal{K,V,X,Y})=\{(f(x_t),y_t,f(x),y)\mid(x,y)\in \mathcal{S}\}
\end{equation}

\subsection{Speech Retrieval}
\label{sec:retrieval}
The datastore is organized as a speech inverted index, which allows us to retrieve similar speech sequences using a term frequency (TF) method similar to text information retrieval. During inference, we use the same speech tokenizer as in the database creation phase and align input speech $\hat{x}$ with the ASR transcription hypothesis to generate the query embedding $f(\hat{x}_t)$ for each token $t$, This process helps us find the token-level k-nearest neighbors (kNN) $N_{k}$. All retrieval results are grouped by the original $f(x)$, denoted as $N_{f(x)}$, to calculate the final sequence level score for $(f(\hat{x}),f(x))$, and each group has $i$ tokens. Specifically, we simply use the following formula to sum the token-level scores for each example:
\begin{equation}
   \text{Score}(f(\hat{x}),f(x))=\sum_{\substack{(k_i,v_i,f(x),y)\in N_{f(x)}}}d(f(\hat{x}_t),k_i)
\end{equation}
where $d(\cdot, \cdot)$ denotes cosine similarity. Finally, we set a threshold filter out examples with low similarity score.

%
RobustGER \cite{robust} shows that in token-aligned N-best lists, error transcription tokens tend to have multiple different values in the same position, while tokens in the same situation tend to be correct transcriptions. We use this information to prune the query sequence, removing the speech tokens in the query that have the same token in the N-best list. The pruning process is illustrated by the red token (C) in Figure \ref{fig:arch}. This allows the LLM to focus only on the erroneous parts, thereby reducing complexity. 

\subsection{LLM Prompt}
As shown in Figure \ref{fig:arch}, after aligning the speech token sequence $f(x)$ with a speech tokenizer, it is fed into a speech adapter to align with the LLM token space and dimensions. Here, we use a feedforward network (FFN) as the adapter. The output of the FFN is given by: $Z = \text{FFN}(f(x))$

We also introduce a model adapter for our LA-RAG task. We employ LoRA \cite{lora} for parameter-efficient fine-tuning, aiming to learn the mapping between the speech token and its correct text token. This enables the LLM to learn the correct text token to the input speech token via ICL during the inference stage. More formally, let $\{Z^0, \cdots, Z^{M-1}\}$ be the FFN output of the top M speech tokens, $\{Y^0, \cdots, Y^{M-1}\}$ be the embedding output of the corresponding text tokens. $\hat{X}$ represents the input speech tokens, with N-best embeddings denoted as $\{\hat{Y}^{0}, \cdots, \hat{Y}^{N-1}\}$. The prompt fed into the LLM can finally be written as:
\begin{equation}
    \text{Concat}(Z^0,Y^0,\cdots,Z^{M-1},Y^{M-1},\hat{X},\hat{Y}^{0},\cdots,\hat{Y}^{N-1})
\end{equation}
Our speech-to-speech retrieval method is a general approach that can be easily generalized to other speech tasks.



\section{Experimental Setup}

\subsection{Dataset}
We utilize both Mandarin and dialect datasets to evaluate the performance of the pre-trained ASR model in sufficiently and insufficiently trained scenarios respectively. The datasets include AISHELL-1 \cite{bu2017aishell1opensourcemandarinspeech} (178 hours, Chinese) and the KeSpeech \cite{kespeech} subdialect datasets. These subdialects encompass Mandarin (589 hours), JiangHuai (46 hours), JiLu (59 hours), ZhongYuan (84 hours), and Southwestern (75 hours).

\subsection{Implementation Details}
We employed the Whisper-Medium model as our base ASR system, and from which we obtained the input and N-best transcriptions. To evaluate different speech tokenization methods, we tested both the CTC and the AED approaches. Specifically, we used the SenseVoice-Small model \cite{sensevoice} for CTC tokenzier and the Whisper-Small model \cite{whisper} for AED tokenzier. Both pre-trained models demonstrated comparable performance on standard open-source ASR test sets. Additionally, for LLM decoding, we adopt LLaMA 3 8B \cite{llama3} from Huggingface. To enhance its performance, a LoRA adapter with a rank of 8 is integrated into each layer of LLaMA. We also implement a simple structured linear projector consisting of two linear layers with an intermediate hidden layer dimension of 2048.

For retrieval, we utilize FAISS \cite{faiss} to retrieve the approximate $k$-nearest neighbors, where $k$ is set to 128. The sequence filter threshold is set to 0.5. For evaluation metrics, we employ the Character Error Rate (CER).

The input to our model comprises the retrieved speech examples mentioned in Section \ref{sec:retrieval}, along with input speech tokens and the 5 best transcripts generated by Whisper. The model is trained for 25 epochs with early stopping to prevent overfitting. We use the Adam optimizer \cite{adam} and experiment with a learning rate of $5 \times 10^{-4}$. Training is conducted on 8 GPUs to leverage efficient parallel processing. An effective batch size of 32 is used, and a weight decay of $1 \times 10^{-2}$ is applied.

\section{Results}
The results of ASR on six datasets, including AISHELL and KeSpeech, are presented in Table \ref{main}, where the training data is used to construct the datastore.

Specifically, HyPoradise refers to \cite{HyPoradise}, which uses the N-best results of the ASR model as LLM prompts for error correction. Whispering LLaMA refers to \cite{whispering_llama}, which contrasts with HyPoradise by adding speech signals to the LLM and achieves more efficient results. Neither method, however, employs retrieval-augmentation to acquire external knowledge.

kNN-CTC, as described in \cite{knn_ctc}, utilizes a external datastore and generally produces better results than the aforementioned methods. However, kNN-CTC uses a small model, lacking the capability of learning similar examples through LLM ICL and finding the optimal token using N-best results. Moreover, according to prior studies \cite{jiang2021learning, speaker_smoth}, such methods is more likely to introduce noise or overfitting during decoding.

Two speech tokenizers were implemented for our LA-RAG. For CTC-based LA-RAG, which constructs the datastore similarly to kNN-CTC as mentioned in Section \ref{sec:datastore}, the lowest CER was achieved among all methods. For AED-based LA-RAG, which employs a different datastore creation method from CTC-based LA-RAG, the average score was similar, with some test sets surpassing the results of CTC-based LA-RAG. Additionally, we observed that our method got more significantly improved performance on accented test sets (max 2.14 CER decrease) than AISHELL and Mandarin. This improvement is attributed to LA-RAG’s ability to help the LLM learn the mapping between pronunciation and correct tokens, which is particularly useful in accent scenarios where the ASR model might not have fully learned the mapping relationships. These experiments demonstrate the effectiveness of our proposed methods.

\begin{table}[]
\centering
\caption{CER (\%) performance with different retrieval settings}
\label{analysis}
\begin{tabular}{ccc}
\hline
Retrieval Type      & JiangHuai &JiLu \\ \hline
Base ASR            & 43.94	&31.61 \\ \hline
Random              & 43.47	&31.4 \\  
Sequence Embedding   & 42.39	&30.81 \\
Text                & 42.72	&31.1 \\
Phoneme             & 42.41	&30.78 \\
No pruning          & 42.04	&30.63 \\ \hline
LA-RAG$_{CTC}$      & 41.8	&30.39 \\ \hline
\end{tabular}
\vspace{-0.5cm} %
\end{table}

\section{Analysis}
\subsection{Retrieval Comparison}

To evaluate the effectiveness of our speech tokenizer of LA-RAG, we compare several related retrieval techniques across two datasets. The results are presented in Table \ref{analysis}.

Firstly, following the methodology in \cite{COSMIC}, we validated the \textbf{Random} sampling approach by selecting the same number of examples from the datastore as our method. While there were some effects, but not very significant. We also compared our method with the use of \textbf{Sequence Embeddings} for kNN speech retrieval by employing the average value of sequence token embeddings, a technique shown to be effective in \cite{wang2024whisperperformspeechbasedincontext}. However, this coarse-grained approach was less effective than our proposed speech token-level retrieval method due to the lower alignment precision required.

Additionally, given the availability of transcription text, we evaluated a simpler and more sophisticated \textbf{Text}-to-text retrieval method. This approach did not perform well on both accent test sets because the transcriptions of accents often contained errors, which limited retrieval accuracy. Furthermore, even with the conversion of text to \textbf{Phonemes}, the improvement was marginal.

Lastly, we assessed the impact of \textbf{No Pruning}, which refers to not removing identical tokens in the N-best list as discussed in Section \ref{sec:retrieval}. The slight increase in CER indicated that the extra tokens that were removed had a detrimental effect. This analysis demonstrates the advantages of our retrieval method, which can be seamlessly extended to other speech-to-speech retrieval tasks, warranting further exploration.

\subsection{Parameter Settings}

\begin{figure}[t]
  \centering
  \includegraphics[width=\linewidth]{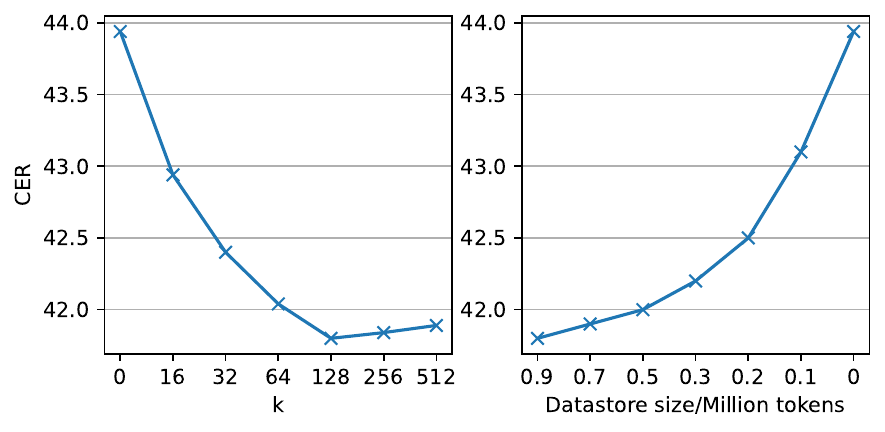}
  \vspace{-0.7cm}
  \caption{Left side is the CER trend when use different top k, right side is the CER trend in different sample datastore size. }
  \label{fig:analysis}
  \vspace{-0.5cm} %
\end{figure}

Figure \ref{fig:analysis} illustrates the impact of varying the top-k parameter and datastore size on performance using the JiangHuai test set and a CTC-based method. Optimal performance was observed at a top-k value of 128. Further increasing the retrieval number led to a performance decline due to noise, though this was mitigated by our threshold control filters described in Section \ref{sec:retrieval}.

The datastore size also influences performance. A larger datastore is preferable as it provides more external knowledge, but it may result in slower retrieval speeds. Given that our datastore currently contains millions of entries, we utilize GPU acceleration through search libraries such as FAISS and employ approximate retrieval methods to ensure the retrieval time remains within 50ms. Addressing the slowdown issue as the datastore grows larger remains a subject for future research.

\section{Conclusion}

In this study, we present a novel RAG paradigm for LLM-based ASR. By leveraging fine-grained speech datastores and precise token-level alignments achieved through pre-trained CTC and AED models, our method significantly enhances LLM-based ASR accuracy, particularly in accent variation scenarios. The experimental results demonstrate consistent improvements across various datasets, including Mandarin and Chinese dialects, with a notable reduction in the CER. This approach highlights the potential for integrating similar speech examples into LLMs and offers a solution for enhancing ASR performance, even under diverse speech conditions. In the future, we plan to generalize our RAG method to other tasks and other languages for speech.


\bibliographystyle{IEEEtran}
\bibliography{template}

\end{document}